# Replicator equations and the principle of minimal production of information

G.P. Karev

Lockheed Martin MSD, National Institute of Health,
Bldg. 38A, Rm. 5N511N, 8600 Rockville Pike, Bethesda, MD 20894, USA
E-mail: *karev@ncbi.nlm.nih.gov*

**Abstract.** Many complex systems in mathematical biology and other areas can be described by the replicator equation. We show that solutions of a wide class of replicator equations minimize the KL–divergence of the initial and current distributions under time-dependent constraints, which, in their turn, can be computed explicitly at every instant due to the system dynamics. Therefore, the Kullback principle of minimum discrimination information, as well as the maximum entropy principle, for systems governed by the replicator equations can be derived from the system dynamics rather than postulated. Applications to the Malthusian inhomogeneous models, global demography, and the Eigen quasispecies equation are given.

## 1. Introduction

The *principle of maximal entropy*, stated most briefly, posits: "when we make inferences based on incomplete information, we should draw them from that probability distribution that has the maximum entropy permitted by the information we do have" [19].

Here "entropy" means the Shannon–Gibbs entropy of a discrete distribution $\{p_i\}$, $S[p] = -\sum_i p_i \log p_i$. S. Kullback [25] formulated a similar (and, formally, a more general) principle using the KL–divergence of the distribution $p$ from $m$. KL–divergence is defined in [26] as

$$I[p:m] = \int_A p(x) \log \frac{p(x)}{m(x)} dx = E_p[\log \frac{p(x)}{m(x)}].$$

The value $S[p:m] = -I[p:m]$ is also known as the relative or cross entropy or information entropy. The KL–divergence allows for unequal prior probabilities $m$ and remains well-defined for continuous distributions, in contrast to the Shannon–Gibbs entropy, which can become undefined for non-discrete probabilities. The KL–divergence is always non-negative but not symmetric, therefore it is not a true distance between distributions.

The inference of $p$ by minimizing $I[p:m]$ (maximizing $S[p:m]$) is known as the *principle of minimum discrimination information*, MinxEnt, which is equivalent to the principle of maximum relative entropy, MaxEnt. Jaynes [18], [19] and his followers have shown that essentially all known statistical mechanics can be derived from the MaxEnt. During the last decades, these methods have been successfully applied to the analysis of a vast number of phenomena.

The rationale of the MaxEnt method is substantially different from that of other statistical methods. According to Jaynes [18] "the probability assignment which most honestly describes what we know should be the most conservative assignment in the sense that it does not permit one to draw any conclusions not warranted by the data". A similar rationale for the MinxEnt method was proposed by Kullback [22]: given new facts, a new distribution $p$ should be chosen, which is as hard to discriminate from the original distribution $m$ as possible; so that the new data produce as small an information gain $I[p:m]$ as possible. These facts, or knowledge, or experimental conditions, or given physical (biological) constraints can typically be expressed as expectation values over the unknown probability $p$. Shore and Johnson [33] suggested an axiomatic explanation of the MaxEnt method as a method for updating probabilities; this approach was developed further by several researchers (e.g., [3], [35]).

In many biological applications the KL–divergence $I[p:m]$ can be interpreted as *production of information* [6]; accordingly, MinxEnt can be reformulated as the *principle of minimal production of biological information*.

Is it possible to derive the principle of maximal entropy from the basic laws and fundamental theories? This problem has been discussed in the literature for a long time. Actually, it was clearly formulated by A. Einstein [8] who argued that the statistics of a system should follow from its dynamics and, in principle, could not be postulated a priori.

A partial solution of this problem is found for systems governed by so called replicator equations. Having in mind models of mathematical biology, we denote $n_i(t)$ the size of the $i$-th species at time $t$, $N = \sum_i n_i$ the total size of the system, $x_i = n_i / N$ the concentration of the $i$-th species in the system, and $F_i$ the per capita growth rate of the individuals of the $i$-th species. Then $\frac{dn_i}{dt} = F_i n_i$, and the concentrations solve the replicator equation

$$\frac{dx_i}{dt} = (F_i - \sum_j x_j F_j) x_i. \tag{1}$$

The replicator equation (RE) is among the basic tools in mathematical ecology, genetics, and mathematical theory of selection and evolution. The finite-dimension RE with $i = 1,...m$ can be considered as a particular case of the Lotka-Volterra equation, but more abstract RE with a set of "species" indexed by elements of a more complex set $A$ are also of use. For instance, the species can be indexed by the reproduction rate, and then $A$ is an interval, or species can be indexed by the birth and death rates, and then $A$ may be a rectangle, or species can consist of all individuals with a fixed set of genes, and then $A$ is a subset of all possible genotypes, that is a set of the sequences formed by a 4-letter alphabet.

During the last decades it has been discovered that similar models (also known as systems with inheritance or selection systems) appear not only in population genetics and selection theory [4, 13, 14] but also in very different areas, such as theoretical ecology and dynamical game theory [16].

The replicator equations naturally arise when mathematical models of systems with intrinsic heterogeneity with respect to some inherited characters are considered; it is assumed that the heterogeneity implies existence of selective differences between individuals. The replicator equations describe the dynamics of the distribution of the characters under selective forces. The changes of this distribution literally mean the production of information, which can be measured with KL-divergence between the initial and current distributions.

In this paper we study the dynamical model of an inhomogeneous population, which we also refer to as a selection system. We show that 1) the solutions a wide class of replicator equations have the form of time-dependent Boltzmann distributions; conversely, every time-dependent Boltzmann distribution satisfies a replicator equation; 2) the distribution of the characters of a selection system solves a replicator equation; conversely, each replicator equation can be associated with a selection system whose distribution solves that replicator equation; 3) the solution of a replicator equation minimizes *at every instant* the KL–divergence of the initial and current distributions (or the production of information) at some natural constraints; these constrains, in their turn, can be computed explicitly at every moment due to the selection system dynamics. The main conclusion follows from these results: the minimal KL-divergence between current and initial distributions is an intrinsic property of the solutions of replicator equations and hence the MinxEnt principle can be derived from the dynamics of the associated selection system instead of being postulated. The obtained results are applied to some particular selection systems, namely, the Malthusian inhomogeneous models, the model of global demography, models of tree stand self-thinning, and the quasispecies theory.

**2. MaxEnt algorithm and the Boltzmann distributions**

The principle of maximum entropy is useful only when applied to *testable information,* i.e. when one can determine whether a given distribution is consistent with it. Informally, we suppose that we can measure only a finite set of traits of interest; as a rule, the testable information is given as the mean of these measurable values. So, assume that expected values of some $n$ variables $\varphi_s$ over the unknown pdf $p$, $E_p[\varphi_s]$, are given:

$$E_p[\varphi_s] = A_s, s = 1,...n. \qquad (2)$$

The variables $\varphi_s(\mathbf{a})$, $\mathbf{a} \in A$, which we will also refer to as traits, are supposed to be defined on a probabilistic space $(A, \mathbf{A}, m)$. The distribution $p^*$ that maximizes the relative entropy $S[p:m]$ subject to the constraints (2) is

$$p^*(\mathbf{a}) = \frac{1}{B}\exp(-\sum_{s=1}^{n}\lambda_s\varphi_s(\mathbf{a}))m(\mathbf{a}). \qquad (3)$$

The normalization factor $B(\boldsymbol{\lambda}) = \int_A \exp(-\sum_{s=1}^{n}\lambda_s\varphi_s)m(\mathbf{a})d\mathbf{a}$ where $\boldsymbol{\lambda} = (\lambda_1,...\lambda_n)$; $B(\boldsymbol{\lambda})$ is known in statistical physics as the partition function and $\exp(-\sum_{s=1}^{n}\lambda_s\varphi_s(\mathbf{a}))$ is the Boltzmann factor. The Lagrange multipliers $\lambda_s$ solve the system

$$-\partial \ln B / \partial \lambda_s = A_s.$$

The maximum value of the cross entropy is equal to $S[p^*:m] = \ln B(\boldsymbol{\lambda}) + \sum_{s=1}^{n}\lambda_s A_s$.

Probability distribution (3) is the least-biased distribution consistent with the available information (2) because, by construction, it contains this information alone. The distribution of the form (3) is often called the Boltzmann distribution. Thus, the equilibrium states of all systems to which the MaxEnt principle can be applied is described by Boltzmann distributions (3).

Let us define the generalized time-dependent Boltzmann distribution

$$P_t(\mathbf{a}) = \frac{\exp(\Phi(t,\mathbf{a}))}{Z(t)}P_0(\mathbf{a}) \qquad (4)$$

where $\Phi(t,\mathbf{a})$ is a smooth functions of time, $P_0(\mathbf{a})$ is a given initial distribution and

$Z(t) = E^0[\exp(\Phi(t,.))]$ is the normalization factor.

Hereinafter we use the notation $E^t[f] = \int_A f(\mathbf{a})P_t(\mathbf{a})d\mathbf{a}$. Remark that $\exp(\Phi)$ and $Z$ are analogues of the Boltzmann factor and the partition function, respectively. The relative entropy of the generalized Boltzmann distribution is equal to

$$S[P_t : P_0] = -E^t[\ln(P_t / P_0)] = \ln Z(t) - E^t[\Phi(t,.)] =$$

$$\ln E^0[\exp(\Phi)] - E^0[\Phi \exp(\Phi)] / E^0[\exp(\Phi)]. \tag{5}$$

Below we prove that the distributions of a wide class of inhomogeneous population models are the generalized Boltzmann distributions (4), which coincide with the distributions computed according to the MaxEnt algorithm at the constraints taken as the current mean values of the traits. It implies that the MaxEnt principle is valid for this class of dynamical systems not only in the equilibrium, but *at each point of the system trajectory*, even when the system is far from equilibrium and even if the system has no equilibrium at all.

### 3. Replicator equations, selection systems and generalized Boltzmann distributions

Instead of the simplest population model and replicator equation (1), let us explore a more general replicator equation

$$dP_t(\mathbf{a})/dt = P_t(\mathbf{a})(F(t,\mathbf{a}) - E^t[F(t,\mathbf{a})]). \tag{6}$$

where $F(t,\mathbf{a})$ is a smooth function of $t$ and a measurable function of $\mathbf{a}$. We show that this equation describes the evolution of the parameter distribution in the "associated" selection system. Let us consider an inhomogeneous population in which every individual is characterized by its own value of the vector-parameter $\mathbf{a} = (a_1,...a_n)$. In general, the parameters $a_i$ may have different origin; the vector $\mathbf{a}$ can be considered as the microstate of the system. Let $l(t,\mathbf{a})$ be the density of individuals in the state $\mathbf{a}$ at the moment $t$, so that $\int_v l(t,\mathbf{a})d\mathbf{a}$ is the total number of individuals having parameter values $\mathbf{a}$ in the phase volume $v$. Let $F(t,\mathbf{a})$ be the (Malthusian) fitness of an individual; in general, it depends on its state **a** and on the "environment" that may changes with time. The associated selection system is defined by the following equations

$$dl(t,\mathbf{a})/dt = l(t,\mathbf{a})F(t,\mathbf{a}), \tag{7}$$

$$P_t(\mathbf{a}) = l(t,\mathbf{a})/N(t)$$

where $N(t) = \int_A l(t,\mathbf{a})d\mathbf{a}$ is the total population size at $t$ instant. The initial distribution $P_0(\mathbf{a})$ and the initial population size $N(0)$ are supposed to be given. The statements collected in the following proposition are actually known in different contexts.

**Proposition 1**.

i) The current pdf $P_t(\mathbf{a})$ of the associated selection system (7) solves the replicator equation (6);

ii) The total population size satisfies the equation $dN/dt = NE^t[F]$;

iii) Replicator equation (6) for a given initial distribution $P_0(\mathbf{a})$ has a unique solution.

Indeed, $\dfrac{dN}{dt} = \dfrac{d}{dt}\int_A l(t,\mathbf{a})d\mathbf{a} = \int_A l(t,\mathbf{a})F(t,\mathbf{a})d\mathbf{a} = N(t)E^t[F];$

$$\frac{d}{dt}P_t(\mathbf{a}) = \frac{d}{dt}\frac{l(t,\mathbf{a})}{N(t)} = \frac{l(t,\mathbf{a})F(t,\mathbf{a})}{N(t)} - \frac{l(t,\mathbf{a})}{N^2(t)}\frac{dN(t)}{dt} = P_t(\mathbf{a})(F(t,\mathbf{a}) - E^t[F(t,\mathbf{a})]).$$

Next, let $P^1_t, P^2_t$ solve the replicator equation and $P^1_0 = P^2_0$. Then

$$\frac{d}{dt}\ln(\frac{P^1_t(\mathbf{a})}{P^2_t(\mathbf{a})}) = 0, \text{ hence } \frac{P^1_t(\mathbf{a})}{P^2_t(\mathbf{a})} = const = \frac{P^1_0(\mathbf{a})}{P^2_0(\mathbf{a})} = 1 \text{ for all } t. \text{ Q.E.D.}$$

If the reproduction rate $F(t,\mathbf{a})$ for model (7) is known explicitly as a function of $t$, then we can define the reproduction coefficient of the selection system for the time interval $[0,t)$ as

$$K^t(\mathbf{a}) = \exp(\Phi(t,\mathbf{a})) \text{ where } \Phi(t,\mathbf{a}) = \int_0^t F(u,\mathbf{a})du. \tag{8}$$

It is easy to check that

$$l(t, \mathbf{a}) = l(0, \mathbf{a}) K^t(\mathbf{a}),$$

$$N(t) = N(0) E^0[K^t], \tag{9}$$

$$P_t(\mathbf{a}) = P_0(\mathbf{a}) K^t(\mathbf{a}) / E^0[K^t]. \tag{10}$$

We have shown that in order to solve the replicator equation one can find the solution of the associated selection system; its current distribution (10) is equal to the desired solution of the replicator equation due to uniqueness. Conversely, if the solution of the replicator equation, the pdf $P_t(\mathbf{a})$, is known, then one can solve the equation $dN/dt = NE^t[F]$ and then obtain the solution of model (7) by the formula $l(t, \mathbf{a}) = P_t(\mathbf{a}) N(t)$. Hence, problems (6) and (7) are equivalent.

Now we can show that the set of all possible solutions of the replicator equations coincides with the set of generalized Boltzmann distributions (4). Let

$$P_t(\mathbf{a}) = \frac{\exp(\Phi(t, \mathbf{a}))}{Z(t)} P_0(\mathbf{a}); \text{ denote } F(t, \mathbf{a}) = \frac{d}{dt} \Phi(t, \mathbf{a}).$$

**Proposition 2**. *Any generalized Boltzmann distribution (4) solves the replicator equation (6). Conversely, if the distribution $P_t(\mathbf{a})$ satisfies the replicator equation, then it is the generalized Boltzmann distribution.*

Indeed, if $P_t(\mathbf{a})$ is of the form (4), then $\dfrac{d \ln P_t(\mathbf{a})}{dt} = \dfrac{d}{dt}(\Phi(t, \mathbf{a}) - \ln Z) = F(t, \mathbf{a}) - \dfrac{1}{Z}\dfrac{dZ}{dt}$, and $\dfrac{1}{Z}\dfrac{dZ}{dt} = \dfrac{E^0[\exp(\Phi) F]}{Z(t)} = E^t[F]$, so $P_t(\mathbf{a})$ solves equation (6).

Conversely, if $P_t(\mathbf{a})$ satisfies equation (6), then it is a distribution of associated system (7) and hence is of the form (10), i.e. $P_t(\mathbf{a})$ is a generalized Boltzmann distribution with the Boltzmann factor equal to the reproduction coefficient for the interval $[0, t)$ of system (7), $K^t(\mathbf{a}) = \exp(\Phi(t, \mathbf{a}))$. Q.E.D.

Proposition 2 shows that the generalized Boltzmann distributions and its dynamics are completely described by the replicator equations. It does not mean, of course, that these distributions can not solve other equations.

Within the framework of the selection model, the rate of production of information is described by the following equation.

**Proposition 3**. The rate of production of information for selection systems (7) satisfies the equation

$$\frac{dI[P_t : P_0]}{dt} = Cov^t[F, \Phi]. \tag{11}$$

This equation follows, after simple algebra, from equalities (4) and (5) but it is instructive to derive it from the 2$^{nd}$, or complete Price equation ([30], see also [29]):

$$dE^t[z]/dt = Cov^t[F, z] + E^t[dz/dt] \tag{12}$$

where $z(t, \mathbf{a})$ is an arbitrary trait. The Price equation is valid at very general conditions (see [32], ch.6); in our case (12) easily follows from (4). Applying the Price equation to $z = \Phi$, we get

$$dE^t[\Phi]/dt = Cov^t[F, \Phi] + E^t[F]. \tag{13}$$

Next, $\dfrac{dI[P_t : P_0]}{dt} = \dfrac{d}{dt}(E^t[\Phi] - \ln E^0[\exp(\Phi)]) = (Cov^t[F, \Phi] + E^t[F]) - \dfrac{E^0[F \exp(\Phi)]}{E^0[\exp(\Phi)]} =$

$= Cov^t[F, \Phi]$ since $\dfrac{E^0[F \exp(\Phi)]}{E^0[\exp(\Phi)]} = E^t[F]$ due to (4). *Q.E.D.*

As a corollary, we obtain a conservation law for selection system (7).

**Proposition 4.** For all $t$

$$I[P_t : P_0] - E^t[\Phi(t,.)] + \ln N(t) = const, \tag{14}$$

and this constant is equal to $\ln N(0)$.

Indeed, subtracting (13) from equation (11), (13) and utilizing that $dN/dt = NE^t[F]$, we obtain

$\dfrac{d}{dt}(I[P_t : P_0] - E^t[\Phi] + \ln N) = 0$, hence

$I[P_t : P_0] - E^t[\Phi(t,.)] + \ln N(t) = \ln N(0)$, because $I[P_0 : P_0] = 0$ and $\Phi(0, \mathbf{a}) = 0$. *Q.E.D.*

## 4. MinxEnt and the solutions of replicator equations

In what follows we will suppose that the reproduction rate per individual, i.e., the individual fitness, can be represented as a finite sum of the form

$$F(t,\mathbf{a}) = \sum_{i=1}^{n} g_i(t)\varphi_i(\mathbf{a}). \tag{15}$$

Rationalization of this supposition is twofold. Mathematically, let us recall that a function of two variables, $f(x,y)$, can be well approximated with finite sums $\sum_i g_i(x)\varphi_i(y)$ under some natural conditions (such as uniform continuity in a finite area). In biological applications, we can consider the individual fitness that depends on a given finite set of traits labeled $i=1,...n$. The function $\varphi_i(\mathbf{a})$ describes quantitative contribution of a particular $i$-th trait to the total fitness, depending on the individual value of the vector-parameter $\mathbf{a}$. For example, $\mathbf{a}$ may be an individual genotype and then $\{\varphi_i(\mathbf{a})\}$ is the set of phenotypical traits of interest. The function $f_i(t)$ describes relative importance of the trait contribution depending on the environment, population size, etc.

The reproduction coefficient (8) for the time interval $[0,t)$ under condition (15) is given by

$$K^t(\mathbf{a}) = \exp(\sum_{i=1}^{n} q_i(t)\varphi_i(\mathbf{a})) \text{ where } q_i(t) = \int_0^t g_i(u)du.$$

It follows from Proposition 1 that the current distribution of the selection system

$$dl(t,\mathbf{a})/dt = l(t,\mathbf{a})F(t,\mathbf{a}) \text{ with } F(t,\mathbf{a}) = \sum_{i=1}^{n} g_i(t)\varphi_i(\mathbf{a}) \tag{16}$$

is the generalized Boltzmann distribution

$$P_t(\mathbf{a}) = \frac{1}{E^0[K^t]}K^t(\mathbf{a})P_0(\mathbf{a}) = \frac{1}{E^0[\exp(\sum_{i=1}^{n} q_i(t)\varphi_i)]}\exp(\sum_{i=1}^{n} q_i(t)\varphi_i(\mathbf{a}))P_0(\mathbf{a}). \tag{17}$$

Condition (15) allows us to define for the associated selection system (7) the Boltzmann factor $\exp(\Phi)$ with $\Phi(\mathbf{q},\mathbf{a}) = \sum_{i=1}^{n} q_i \varphi_i(\mathbf{a})$ and the partition function

$$Z(\mathbf{q}) = E^0[\exp(\sum_{i=1}^{n} q_i \varphi_i)] \tag{18}$$

where $\mathbf{q}(t) = (q_1(t),...q_n(t))$.

Remark, that the partition function (18) has a clear biological sense within the frameworks of selection system (7), (15): $Z(\mathbf{q}) = E^0[K^t]$ is proportional to the ratio of the current and initial population sizes due to formula (9).

Let us explore the properties of the solution (17) of replicator equation (6), (15) and associated selection system (16). Denoting $\vartheta = (\varphi_1,...\varphi_n)$ let $p_t(\vartheta)$ be the pdf of the random vector $\vartheta$ at $t$ moment, i.e. $p_t(x_1,...x_n) = P_t(\varphi_1 = x_1,...\varphi_n = x_n)$. Let $\boldsymbol{\lambda} = (\lambda_1,...\lambda_n)$; denote

$$M(\boldsymbol{\lambda}) = \int_A \exp(\sum_{i=1}^{n} \lambda_i \varphi_i(\mathbf{a})) P_0(\mathbf{a}) d\mathbf{a} = \int_R \exp(\sum_{i=1}^{n} \lambda_i x_i) p_0(x_1,...x_n) dx_1...dx_n$$

the moment generation function (mgf) of the initial distribution. It is well known that mgf uniquely determines the distribution. The mgf-s of all widely used distributions (such as normal, exponential, Gamma-distribution, etc.) are known in the analytical form. In general, one can consider the mgf of any given initial pdf as a known or at least as easily computable function. What is important is that the partition function (18) for the associated selection system is readily computed if the mgf of the initial system distribution is given: $Z(\mathbf{q}) = M(\mathbf{q})$.

Now we are able to formulate the main results.

**Theorem.**

1) *Let $P_t$ be the solution* (17) *of replicator equation* (6), (15). *Then at every moment $t$ the distribution $P_t$ provides minimum of $I[P_t : P_0]$ over all probability distributions compatible with the constraints $A_i(t) = E^t[\varphi_i]$, $i = 1,...n$;*

2) *The values of constraints evolve due to the associated selection system and at each time moment are equal to*

$$A_i(t) = E^0[\varphi_i K^t] / E^0[K^t] = \partial_i \ln(M(\mathbf{q}(t))); \qquad (19)$$

3) *Dynamics of the constraints are determined by the covariance equation*

$$dA_i(t)/dt = Cov^t[F, \varphi_i]; \qquad (20)$$

*the current covariance $Cov^t[\varphi_i, \varphi_k]$ of the traits $\varphi_i, \varphi_k$ can be computed by the formula*

$$Cov^t[\varphi_i, \varphi_k] = [\partial^2_{ik} M(\mathbf{q}(t)) - \partial_i M(\mathbf{q}(t)) \partial_k M(\mathbf{q}(t))] / M(\mathbf{q}(t)).$$

Assertion 1) can be proven directly by solving the corresponding variation problem but one can use known results, see s.2. Let us compare the MaxEnt distribution $p^*$ and the solution of the replicator equation at moment $t$:

$$p^*(\mathbf{a}) = \frac{1}{B} \exp(-\sum_{i=1}^n \lambda_i \varphi_i(\mathbf{a})) m(\mathbf{a}), \quad B(\boldsymbol{\lambda}) = E^m[\exp(-\sum_{i=1}^n \lambda_i \varphi_i)]$$

where $\boldsymbol{\lambda} = \{\lambda_i\}$ is the solution of the system $-\partial_i \ln B = A_i$, and

$$P_t(\mathbf{a}) = \frac{1}{E^0[K^t]} \exp(\sum_{i=1}^n q_i(t) \varphi_i(\mathbf{a})) P_0(\mathbf{a}), \quad E^0[K^t] = E^0[\exp(\sum_{i=1}^n q_i(t) \varphi_i)]$$

where $q_i(t) = \int_0^t g_i(r) dr$. Identifying the pdfs $P_0$ and $m$, we see that then $B(\boldsymbol{\lambda}) = M(-\boldsymbol{\lambda})$, and the pdf $P_t$ at given instant $t$ coincides with $p^*$ if $\lambda_i = -q_i(t)$. We have already proven that if the constraints are defined as $A_i(t) = E^0[\varphi_i K^t] / E^0[K^t]$ then $A_i(t) = E^t[\varphi_i]$ and $\{q_i(t)\}$ solve the system $\partial_i \ln(M(\mathbf{q}(t))) = A_i(t)$. This system is identical to that which defines the Lagrange multipliers, $-\partial_i \ln Z(\boldsymbol{\lambda}) = A_i$. Hence, the last system has the solution $\lambda_i = -q_i(t)$, and the MaxEnt distribution $p^*$ under constraints $A_i(t)$ exists and coincides with the solution $P_t$ of the replicator equation.

Next, the first equality in (19) directly follows from (17) and the second one follows from the definition of mgf $M(t)$. The equality (20) follows from the Price equation (12) as $d\varphi_i / dt = 0$.

Now let us collect together some useful formulas.

**Proposition 5**. *The production of information $I[P_t : P_0]$ can be computed with the help of the following formulas:*

$$I[P_t : P_0] = E^t[\Phi(t,.)] - \ln(N(t)/N(0));$$

$$I[P_t : P_0] = E^0[\Phi e^\Phi] / E^0[e^\Phi] - \ln E^0[e^\Phi];$$

$$I[P_t : P_0] = \sum_{i=1}^{n} q_i(t)\partial_{q_i} \ln(M(\mathbf{q}(t))) - \ln(M(\mathbf{q}(t))).$$

The theory developed above for selection systems with the fitness of the form (15) can be applied immediately only if the time-dependent components $g_i(t)$ are known explicitly. As a rule, it is not the case for most interesting and realistic models where the time-dependent components should be computed depending on the current population characteristics. For example, a well-known logistic model corresponds to the function $g(t) = 1 - N(t)/B$ where $B = const$ is the upper boundary of the population size; Allee-type models, which also take into account the lower boundary $b$ of the population density, use the function $g(t) = (1 - N(t)/B)(N(t)/b - 1)$.

Suppose that the individual reproduction rate can depend on some integral characteristics of the system, which we call "regulators", having the form $H(t) = E^t[h]$ or $S(t) = N(t)E^t[s]$ where $h, s$ are given functions. The total system size $N(t)$ is also a regulator of a special importance. Suppose also that the fitness of every individual is determined by a given set of traits and may depend on the total population only through the regulators. In such a model with *self-regulated fitness* the regulators and hence the reproduction rate are not given as explicit functions of time but should be computed together with the current pdf $P(t, \mathbf{a})$ at each time moment.

It was proved in [24] that a self-regulated selection system can be reduced to an equivalent system of ordinary differential equations (ODEs). These results allow us to define and compute the total population size, the current distribution of the system and the values of all regulators at any time moment. Eventually, all results of this section can be applied to self-regulated selection systems.

## 5. Applications and examples

Dynamics of any inhomogeneous biological system that is not in equilibrium is accompanied by the change of distributions of some or all of its characteristics and hence by the production of information. Let us trace this process in some examples of dynamical models of biological populations.

### 5.1. *Inhomogeneous Malthusian model*

Let $F = \varphi(\mathbf{a})$; we can consider the value $\varphi(\mathbf{a}) = a$ as the distributed parameter and study the simplest replicator equation $dP_t(a)/dt = P_t(a)(a - E^t[a])$. The corresponding inhomogeneous Malthusian model is

$$dl(t,a)/dt = al(t,a). \qquad (21)$$

Let $M(\lambda) = \int_A \exp(\lambda a) P_0(a) da$ be the mgf of the initial distribution. Then the solution of (21) is given by $l(t,a) = \exp(at) l(0,a)$, $N(t) = N_0 M(t)$, and the solution of the replicator equation

$$P_t(a) = P_0(a) \exp(at) / M(t). \qquad (22)$$

We can see that even the simplest replicator equation possesses a variety of solutions depending on the initial distribution. Recall that according to the theorem of S. Bernstein (see, e.g., [9], ch.13.4), a function $M(\lambda)$ is the mgf for some pdf if and only if it is absolutely monotone and $M(0) = 1$. So, the total size of inhomogeneous Malthusian population can change as arbitrary absolutely monotone function $M(t)$ at corresponding initial distribution. According to Proposition 2, $dN/dt = NE^t[a]$ and

$dE^t[a]/dt = Var^t[a] > 0$ (it is the simplest version of the Fisher Fundamental theorem of selection [10], see also [12]); hence, any inhomogeneous Malthusian population increases *hyper-exponentially*.

Next, for these models $F = a$, $\Phi = at$, and according to Proposition 3

$$\frac{dI[P_t : P_0]}{dt} = Cov^t[\Phi, F] = tVar^t[a] \tag{23}$$

so that the production of information increase monotonically and faster then a linear function of time.

The following formula, which connects the relative entropy with the current total size and the mean reproduction rate (see (14)) is also of interest:

$$I[P_t : P_0] = tE^t[a] - \ln(N(t)/N(0)). \tag{24}$$

For practical computations of $I[P_t : P_0]$ at different initial distributions it is convenient to rewrite this formula as

$$I[P_t : P_0] = td(\ln M(t))/dt - \ln M(t). \tag{25}$$

The current distribution of the inhomogeneous Malthusian model provides the minimal production of information at a single constraint, $E^t[a] = A(t)$; this constraint varies with the time due to the model dynamics and can be computed by the formula

$$E^t[a] = \frac{d \ln M(t)}{dt}.$$

Let, for example, parameter $a$ be normally distributed at the initial instant, so that $M(\lambda) = \exp(\lambda^2 \sigma^2 / 2 + \lambda m)$ where $m$ is the mean and $\sigma^2$ is the variance. It is easy to show [22], [23] that the parameter distribution at any $t$ is also normal with the mean $E^t[a] = m + t\sigma^2$ and with the same variance $\sigma^2$. The production of information is equal to $I[P_t : P_0] = t^2 \sigma^2 / 2$, and $I[P_t : P_0] \to \infty$ at $t \to \infty$.

Let parameter $a$ be Gamma-distributed with coefficients $s, k, b$ at the initial instant, so that $M(\lambda) = \exp(\lambda b)(1 - \lambda/s)^{-k}$ for $\lambda < s$. Then the parameter is also Gamma-distributed with coefficients $s - t, k, b$ at the moment $t < s$ [22]. The production of information is $I[P_t : P_0] = k \ln(1 - t/s) + kt/(s - t)$, and $I[P_t : P_0] \to \infty$ at $t \to s < \infty$.

In particular, if $k=1, b=0$, i.e. the initial distribution is exponential with the mean $s$, then $M(\lambda) = (1-\lambda/s)$, $N(t) = N(0)(1-t/s)$ and $I[P_t : P_0] = \ln(1-t/s) + t/(s-t)$.

The latter example is applicable to problems of global demography and early biological evolution, see sections 5.2 and 5.3 below.

## 5.2. *Global demography*

The growth of the world population up to $\sim 1990$ was described with high accuracy by the hyperbolic law $N(t) = C/(T-t)$ with $C \approx 2*10^{11}$ that predicts a demographic explosion at the time $T \approx 2025$ [11]. This formula solves the quadratic growth model $dN/dt = N^2/C$, in which the individual reproduction rate is proportional to the total human population. Apparently, this relationship makes no "biological" sense and cannot be the basis of any realistic theory.

If the mean reproduction rate is the only quantity we can estimate from historical demographic data, then the most likely (the maximum entropy) distribution of the reproduction rate is the exponential one with the estimated mean (see, e.g., [20], s.3.2.1). The above results show (see [22] for details) that the hyperbola $N(t) = C/(T-t)$ is implied not only by the quadratic growth model but also by the more plausible Malthusian inhomogeneous model with an exponentially distributed reproduction rate such that $s = T$ and the mean $E^0[a] = 1/T$.

Given that any real population is inhomogeneous, the simplest inhomogeneous Malthusian model is more acceptable as a starting point for global demography modeling then the quadratic growth model. The population increases in such a way that the distribution of the reproduction rate is exponential at every instant $t < T$ with the mean $E^t[a] = 1/(T-t)$, providing minimum of the production of information $I[P_t : P_0]$ under the constraint $A_t = 1/(T-t)$.

The "demographic explosion" occurs at the moment $t = T$ when not only $N(t) = \infty$, but also $E^t[a] = \infty, Var^t[a] = \infty$ and $I[P_t : P_0] = \infty$. It is a corollary of the obviously unrealistic assumption (incorporated implicitly into quadratic growth model) that the individual reproduction rate may take unlimitedly large values with non-zero probabilities.

When the reproduction rate in the model is bounded, $a \in (0, c)$ and the mean value of the reproduction rate is again prescribed, then according to the MaxEnt principle the initial distribution is the truncated exponential in that interval ([20], s.3.3.1); specifically for real demography data, $c \approx 0.114$ [22]. The result is that $N(t)$ is finite, even though indefinitely increasing, for all $t$, and is very close to the hyperbola for a long time (up to 1990 at corresponding values of coefficients).

As shown previously in [22], the subsequent transition from the Malthusian model to the inhomogeneous logistic model shows a transition from prolonged hyperbolical growth (the phase of "hyper-exponential" development) to the brief transitional phase of "almost exponential" growth accomplished by a sharp increase of the variance of the reproduction rate and, subsequently, to stabilization. We conclude that the hyperbolic growth of the humankind was not an exclusive phenomenon but obeyed the same laws as any heterogeneous biological population. In particular, the minimum of the production of information, i.e. the minimum of the KL-distance between the initial and current distributions of the reproduction rate is achieved at every time moment under the given mean rate at this moment.

### 5.3. *The model of early biological evolution*

Non-homogeneous Malthusian dynamics together with the principle of limiting factors were used in a model of early biological evolution [37]. Each organism was characterized by the vector **a** where the component $a_i$ is the thermodynamic probability that protein $i$ is in its native conformation. The authors suppose that the organism death rate $d$ depends on the stability of its proteins as $d = d_0(1 - \min a_i)$, $d_0$=const. Neglecting possible mutations, the model can be formalized as the selection system $dl(t,\mathbf{a})/dt = l(t,\mathbf{a})B(m(\mathbf{a}) - a_0))$. Here $m(\mathbf{a}) = \min[a_1,...a_n]$, $B = b/(1-a_0)$, $b$ is the birth rate, $a_0$ is the native state probability of a protein; we let $B = 1$. Following [37] we can consider $a_i$ as independent random variables with common pdf $f(a)$. Then the model can be reduced (see [24]) to the inhomogeneous Malthusian equation

$dl(t,m)/dt = l(t,m)(m - a_0)$ with the initial pdf of $m$ $g(m) = n(1 - G(m))^{n-1} f(m)$ where $G(m) = \int_0^m f(a)da$.

This equation can be solved explicitly for given $f(a)$ as it was described in s.5.1. Let $P(t,m)$ be the pdf of $m$ at $t$ moment. Let $f(a) = \exp(-a/T)/T$, $0 < a < \infty$ be the Boltzmann distribution with the mean $T$. Then (see [24]) $P(0,m) \equiv g(m) = n/T \exp(-mn/T)$ and $P(t,m) = (n/T - t)\exp(-m(n/T - t))$ are also the Boltzmann distributions. The total population size $N(t) = N(0)\exp(-a_0 t)\dfrac{1}{1 - tT/n}$, the mean value $E^t[m] = \dfrac{1}{1 - tT/n}$.

The production of information in this model $I[P_t : P_0] = \ln(1 - tT/n) + \dfrac{tT/n}{1 - tT/n}$.

We can see now the similar phenomenon as in the previews example: the population „blows up" at the moment $t_{max} = n/T$, i.e. $N(t)$ and $E^t[m]$ tend to infinity at $t \to t_{max}$. The production of information increases monotonically and sharply tends to infinity at $t \to t_{max}$. So, the classical Boltzmann distribution, which allows arbitrary large values of the parameter $a$ with nonzero probability, has no biological sense within the framework of the Malthusian inhomogeneous model. This problem can be eliminated by taking the "truncated" Boltzmann distribution, which allows only bounded values of the parameter $a$; see [24] for detail.

**5.4.** *Models of tree stand self-thinning*

Recently Dewar and Porte [5] showed that the Principle of maximal relative entropy can be used to explain and predict species abundance patterns in ecological communities in terms of the most probable behavior under given environmental constraints. Here we consider a particular ecological problem of dynamics of tree number in a forest, which is one of the oldest and most important problems in forest ecology. A number of tree interactions, variations in genetic structure, and various environment conditions affect the growth and death of trees in complex ways. It seems to be impossible to take into account all impacts on the death rate of trees in explicit form within the frameworks of a unique model.

A promising way to overcome these difficulties is constructing tree population models with distributed values of the mortality rate $a$. It was shown in [21] that different formulas of forest stand self-thinning can be considered as solutions of inhomogeneous Malthus extinction model

$$dl(t,a)/dt = -acl(t,a), \quad N(t) = \int_A l(t,a)da \qquad (26)$$

with corresponding initial distribution of the mortality rate $a$. Here $c = const$ is a time-scaling parameter.

Let us consider, for example, a known formula suggested by Hilmi (see [21] for references) for tree number $N(t)$ dependently on the age $t$ of tree stand,

$$N(t) = N(0)\exp(a_0(e^{-ct} - 1)). \qquad (27)$$

One can notice that this formula practically coincides with a well known Hompertz function. In what follows we let $c = 1$ for simplicity. It was shown in [21] that (27) is an exact solution of (26) if the mortality rate $a$ has the Poisson distribution with average $a_0$ in an initial instant. Then the distribution of $a$ at any moment of time is also Poisson with the mean $E^t[a] = a_0\exp(-t)$. Indeed, the mgf of the Poisson distribution $M(\lambda) = \exp(a_0(e^{-t} - 1))$, and all assertions follow from formulas of s.5.1.

The production of information in this model can be computed according to the formula (24), $I[P_t : P_0] = -tE^t[a] - \ln(N(t)/N(0)) = a_0(1 - e^{-t}(t+1))$. Hence, the production of information increase with time and tends to a limit value $a_0$ when the number of trees tends to its limit value $N_\infty = N(0)\exp(a_0)$.

5.5. *Quasispecies equation and linear systems*

Quasispecies theory, as it was formulated by Eigen and coauthors [6], [7] is based on the concepts of information theory and studies the equation, which can be written in the form

$$dx(t,i)/dt = \sum_k w_{ik} x(t,k) - w(t)x(t,i). \qquad (28)$$

Here $i$ is a running index attributed to all distinguishable self-reproductive molecular units (sequences) and $x(t,i)$ is the respective concentration; $w_{ij} = A_j Q_{ij}$ is the product of the replication rate (fitness) $A_j$ of sequence $j$ and the mutation probability $Q_{ij}$ from sequence $j$ to $i$, and $w(t) = \sum_i \sum_j w_{ij} x_j(t)$ is the total production of new sequences.

Let us consider the associated inhomogeneous population model

$$dl(t,i)/dt = \sum_k w_{ik} l(t,k). \tag{29}$$

According to Proposition 1, i), in order to solve quasispecies equation (28) we can solve (a simpler) associated system (29) and then find the concentrations $x(t,i)$ by formulas $x(t,i) = l(t,i)/N(t)$ where $N(t) = \sum_i l(t,i)$.

Note that, although quasispecies equation (28) is non-linear, the associated system (29) is a general linear system with constant coefficients for which the solution and the asymptotic behavior are well known. The solution of the quasispecies equation was given in [36] under the following condition: the matrix $W = \{w_{ik}\}$ is diagonalizable, i.e. there exists matrix $V$ such that $V^{-1}WV = D$ where $D$ is a diagonal matrix.

This condition implies that the matrix $W$ has $n$ linearly independent (right) eigenvectors $\mathbf{v}^{(j)}$, $j = 1,...n$ and the columns of the matrix $V$ are these $n$ eigenvectors; the corresponding eigenvalues $\delta_i$ are the elements of the main diagonal of the matrix $D$ (these assertions come from textbooks, see, e.g., [17], ch.1). Under the same condition, let us change the variables

$$z(t,i) = \sum_j h^i{}_j l(t,j) \tag{30}$$

where $h^i{}_j$ are the elements of the matrix $H = V^{-1}$. Remind, that the rows of matrix $H$ are proportional to the left eigenvectors $\mathbf{h}^{(i)}$, $i = 1,...n$ of the matrix $W$. Then

$$dz(t,i)/dt = \sum_{j,k} h^i{}_j w_{jk} l(t,k) = \delta_i z(t,i). \tag{31}$$

The variables $z(t,i)$ are "quasispecies", a certain mixture of the initial species $l(t,j)$. The probabilities $P_t(i) = z(t,i)/\sum_k z(t,k)$ solve the simple replicator equation (see [6])

$$dP_t(i)/dt = P_t(i)(\delta_i - \sum_k \delta_k P_t(k)) \tag{32}$$

The associated linear model (29) is reduced to the Malthusian inhomogeneous model (31) with the Malthusian parameter taking only a finite number of values.

Equation (32) can be easily solved directly; coming back to the initial variables, we can get the solutions of equations (29) and (28) as described previously in [33], but we rather interested in the dynamics of quasispecies distribution and corresponding production of information. Let $M(\lambda) = \sum_i \exp(\lambda \delta_i) P_0(i)$ be the mgf of the initial distribution. Then, according to formula (22),

$$P_t(i) = \frac{P_0(i) \exp(\delta_i t)}{M(t)}. \tag{33}$$

The production of information is equal to

$$I[P_t : P_0] = tE^t[\delta] - \ln M(t) = t d(\ln M(t))/dt - \ln M(t).$$

For example, if the initial distribution of quasispecies is uniform, $P_i(0) = 1/N$, then

$$I[P_t : P_0] = \ln N - \ln \sum_i \exp(t\delta_i) + t \frac{\sum_i \delta_i \exp(t\delta_i)}{\sum_i \exp(t\delta_i)}.$$

A more realistic supposition is that the distribution of the initial species, $p(j) = l(0, j)/N(0)$ was uniform at the initial time moment. Then we can compute the initial distribution of quasispecies using the formula $z(0,i) = \sum_j h^i{}_j l(0, j)$, see (30).

Hence, if the initial distribution of species is uniform, then $z(0,i)$ is proportional to the sum of components of $i$-th left eigenvector $\mathbf{h}^{(i)}$. In a general case, if $p(j)$ is the initial frequency of $j$-th species, then the initial frequency of $i$-th quasispecies is proportional to $E_p[\mathbf{h}^{(i)}]$.

For any initial distribution of species and corresponding distribution $P_0$ of quasispecies, the current distribution of quasispecies has the form of generalized Boltzmann distribution (33). It minimizes the KL-divergence between the current and initial distributions, i.e., the production of information among all distributions compatible with the constraint

$$\sum_i \delta_i P_t(i) = A(t). \tag{34}$$

The constraint (34) is hard to rationalize in terms of the initial model (28) as it prescribes the mean value of eigenvalues of the matrix $W$, but this constraint is quite natural when selection system (31) is under consideration. In this case it has a sense of prescribed mean reproduction rate of the population of quasispecies.

The constraint value at each time moment can be easily computed if the mgf of the initial distribution of quasispecies is known: $A(t) = d \ln(M(t))/dt$.

Let us emphasize that the constrained value is not a constant but evolves according to the covariance equation (20), $dA(t)/dt = Var^t[\delta] = \sum_i (A(t) - \delta_i)^2 P_t(i)$.

The mean reproduction rate of quasispecies at $t$ moment is the unique fundamental constraint, which completely defines the distribution of quasispecies due to the MinxEnt principle.

## 6. Discussion and conclusion

The Principal of maximum relative entropy as a method for inference of unknown distribution was successfully applied for the last decades to the analysis of different physical and statistical problems (see, e.g., Proceedings of 1-27 International Workshops on Bayesian Inference and Maximum Entropy Methods in Science and Engineering). Applications of the notion of relative entropy or information entropy and the MaxEnt principle to various problems of mathematical biology, in particular to genetic models of natural selection ([13]), ecological models ([1], [5], [28], [31]), genomics ([27]) and to replicator dynamics ([2]) also have a long history. In this paper we show that, for a wide class of biological models, selection systems, the dynamical version of MaxEnt principle (in the form of minimal information production) can be derived from the system dynamics instead of being postulated.

The selection system describes a closed population of individuals each of which is characterized by a set of qualitative traits; the values of these traits determine the reproduction rate of the individual. It is supposed that the mean values of the traits are the only information that is known from measurements. The dynamics of the joint distribution of these traits depending on the initial distribution and on correlations between the traits is the main problem of interest, which can be solved effectively. The evolution of distributions of selection systems is governed by a certain class of replicator equations; similar equations appear in different scientific areas. We show that the solution of any replicator equation from this class is a generalized Boltzmann distribution. Having the solution of the replicator equation we can compute the current mean values of the traits at any instant. Then, considering these mean values as constraints, we show that the "MaxEnt distribution", which provides the minimum of the KL – divergence between the initial and current distributions coincides with the solution of the replicator equation. This solution was obtained independently of the MaxEnt algorithm. Hence, the principle of minimum of the production of information, equivalently, the MaxEnt principle, can be considered as the variational principle which governs the selection system dynamics.

There exists an "observer-dependent" view of the entropy and cross-entropy concepts (defended by Jaynes [18], [19] and, subsequently, by many other authors). Briefly, the authors claimed that entropy is a property of our description of a system rather than a property of a system. We show that, at least within the framework of selection systems, we cannot choose whether or not to prescribe the property of minimization of the KL-divergence to the selection systems whose distribution is governed by the replicator equations. It is an intrinsic property of any solution of the replicator equations that is fulfilled due to the system dynamics at any instant of its evolution. We are therefore compelled to adopt the "objective" view of the relative entropy concept and its maximization, at least when the replicator equation is taken as the "basic law". "Subjective -dependent" is only the choice of traits that characterize the system, whose joined distribution is of our interest under condition that we have "testable information" about the traits.

Our approach is illustrated for the Malthusian-like selection systems. As a results of the selection process, the production of information in such systems increase with time being minimal at each time moment over all distributions of the Malthusian parameter compatible with the current values of constrains. We explore some particular Malthusian selection systems which are of considerable intrinsic interest, namely, the model of global demography, the model of early biological evolution, the ecological model of forest self-thinning, and the quasispecies equation. In all cases the mean value of the reproduction rate is considered as the only testable information about the systems.

We show that the standard exponential (Boltzmann) distribution cannot be taken as the initial distribution of the reproduction rate for the Malthusian selection system, because the system "blows up" at certain time instant. The demography model shows the hyperbolic growth discovered by Forster and coworkers. Similar problem appears in the model of early biological evolution. The problem can be eliminated if the initial distribution is truncated exponential, which allows only bounded values of the Malthusian parameter. Considering the quasispecies equation we concentrated on the problem of dynamics of the distribution of a quasispecies system and corresponding production of information. The principal new finding is that the current distribution of quasispecies minimizes the production of information at any initial distribution in any instant. The obtained results can be extended to models of biological populations and communities whose growth is governed by self-regulation processes.

**Acknowledgement.** The author thanks Dr. E. Koonin and Dr. A. Novozhilov and two anonymous reviewers for valuable comments.